\pdfoutput=1

\documentclass[12pt]{scrartcl}

\usepackage[square,sort,comma,numbers]{natbib}
\usepackage{graphicx,color,mathtools,amsfonts,xcolor}
\graphicspath{{../Figures/}}

\usepackage{mathptmx}      

\usepackage{hyperref} 
\hypersetup{linkbordercolor=green}

\begin{document}

\title{Quasi-steady-state approximations derived from the stochastic  model of  enzyme kinetics 
}

\author{Hye-Won Kang\footnote{Department of Mathematics and Statistics, 
		University of Maryland Baltimore County, USA, 
		email:\href{mailto:hwkang@umbc.edu}{hwkang@umbc.edu}  } \hfill  \\
	Wasiur R. KhudaBukhsh\footnote{Department of Electrical Engineering and Information Technology, 
		Technische Universit\"{a}t Darmstadt, Germany, 
	email:\href{mailto:wasiur.khudabukhsh@bcs.tu-darmstadt.de}{wasiur.khudabukhsh@bcs.tu-darmstadt.de}  }      \hspace{1.5mm}\href{https://orcid.org/0000-0003-1803-0470}{\includegraphics[width=3mm]{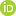}} \hfill 	 \\
	Heinz Koeppl\footnote{Department of Electrical Engineering and Information Technology, 
		Technische Universit\"{a}t Darmstadt, Germany, 
		email:\href{mailto:heinz.koeppl@bcs.tu-darmstadt.de}{heinz.koeppl@bcs.tu-darmstadt.de} } \hfill \\
	Grzegorz A. Rempa{\l}a\footnote{Division of Biostatistics and Mathematical Biosciences Institute, 
		The Ohio State University, USA, 
		email:\href{mailto:rempala.3@osu.edu}{rempala.3@osu.edu}  }     \hspace{1.5mm}\href{https://orcid.org/0000-0002-6307-4555}{\includegraphics[width=3mm]{Figures/ORCID-iD_icon-16x16}} \hfill 
}

\maketitle

\begin{abstract}
	In this paper we derive several quasi steady-state approximations (QSSAs) to  the stochastic reaction network describing  the Michaelis-Menten enzyme kinetics. We show how   the different assumptions about chemical species abundance and reaction rates lead to the standard QSSA (sQSSA), the total QSSA (tQSSA), and the reverse QSSA (rQSSA) approximations. These three QSSAs have  been widely studied in the literature in deterministic ordinary differential equation (ODE)  settings  and several sets of conditions for their validity have been proposed.  By using multiscaling techniques  introduced in  \cite{Kang:2013:STM,Ball:2006:AAM} we   show that  these  conditions  for deterministic QSSAs largely agree with the ones  for  QSSAs  in   the large volume limits of the underlying stochastic enzyme kinetic network. 
	
\end{abstract}


\setcounter{equation}{0}
\nopagebreak
\section{Introduction} \label{intro}
In chemistry and biology, we often come across chemical reaction networks where one or
more of the species exhibit a different intrinsic time scale and tend to reach an equilibrium state quicker than others. Quasi steady state approximation (QSSA) is a commonly used tool to simplify the description of the dynamics of such systems. In particular, QSSA has been widely applied to the important class of reaction networks known as the Michaelis-Menten  models of enzyme kinetics~\cite{Cornish-Bowden:2004:FEK,Segel:1975:EK,Hammes:2012:ECR}.

Traditionally the  enzyme kinetics has been studied using systems of ordinary differential equations (ODEs). The ODE approach allows one to analyze various aspects of the  enzyme dynamics    such as asymptotic  stability. However, it ignores the fluctuations of the enzyme reaction network due to  intrinsic noise and instead focuses on  the averaged dynamics.  If  accounting for this   intrinsic noise is required, the  use of an alternative   stochastic reaction network  approach  may be more appropriate,   especially when some of the species have low copy numbers or when one is  interested in predicting the molecular fluctuations of the system. It is well-known  that such molecular fluctuations in the species with small numbers, and stochasticity in general,  can lead to  interesting dynamics. For instance, in a  recent paper \cite{Perez:2016:MCB}, Perez et al.   gave an account of how intrinsic noise controls and alters the dynamics, and steady state of morphogen-controlled bistable genetic switches. Stochastic models have been strongly advocated by  many in recent literature~\cite{Bressloff:2017:SSB,Assaf:2017:WKB,Newby:2015:BSA,Biancalani:2015:GTS,Bressloff:2013:SNN, Newby:2012:IIN}. In this paper, we consider such stochastic models in the context of QSSA and the Michaelis-Menten enzyme kinetics and relate them   to the  deterministic ones that are well-known  from  the  chemical physics literature.

The  QSSAs are very useful from a practical perspective. They not only reduce the model complexity, but also allow us to better relate it to experimental measurements  by  averaging out the unobservable or difficult-to-measure species. A substantial body of work has been published to justify such QSSA  reductions in deterministic models, typically by means of perturbation theory~\cite{Segel:1989:QSS,Schneider:2000:MRE,Stiefenhofer:1998:QSS,Dingee:2008:NPS,Schnell:1997:CST,Bersani:2011:AEE}. In contrast to this approach,  we derive here the QSSA reductions using stochastic multiscaling techniques \cite{Kang:2013:STM,Ball:2006:AAM}.  Although our approach is applicable more generally,  we focus below on  the three well established enzyme kinetics QSSAs, namely the standard QSSA (sQSSA), the total QSSA (tQSSA), and the reverse QSSA (rQSSA) for the Michaelis-Menten enzyme kinetics. We show that these QSSAs are a consequence of the law of large numbers for the stochastic reaction network under different scaling regimes. A similar approach has been recently taken in~\cite{Kim:2017:RSB} with respect to   a particular type of QSSA (tQSSA, see below in Section~\ref{sec:mm_qssa}). However, our current derivation is different in that it entirely avoids a spatial averaging argument used in \cite{Kim:2017:RSB}. Such an argument requires additional assumptions that are difficult to verify in practice.

The paper is organized as follows.  We first review the Michaelis-Menten enzyme kinetics in the deterministic setting and the discuss corresponding QSSAs in Section~\ref{sec:mm_qssa}. The alternative  model in the stochastic setting  is  introduced in Section~\ref{sec:mm_approximation}, where we also briefly describe  the multiscale approximation technique proposed in \cite{Kang:2013:STM}. Following this, we derive the Michaelis-Menten deterministic sQSSA, the tQSSA and the rQSSA  approximations from the stochastic model analysis in the Sections~\ref{sec:sQSSA}, \ref{sec:tQSSA}, and \ref{sec:rQSSA} respectively. We conclude the paper with a short discussion in Section~\ref{sec:discussion}.

\section{QSSAs for deterministic Michaelis-Menten  kinetics}
 \label{sec:mm_qssa}
The Michaelis-Menten enzyme-catalyzed reaction networks have been studied in depth over past several  decades \cite{Cornish-Bowden:2004:FEK,Segel:1975:EK,Hammes:2012:ECR} and  have been described in various forms.  Although the methods discussed below certainly apply to more general networks  of  reactions describing enzyme kinetics,  in this paper, we adopt the simplest (and minimal) description  for illustration purpose.   In its simplest form,  the  Michaelis-Menten enzyme-catalyzed network of reactions  describes
 reversible binding of a free enzyme ($E$) and a substrate ($S$) into  an enzyme-substrate complex ($C$), and irreversible conversion of the complex to the product ($P$) and the free enzyme. The enzyme-catalyzed reactions are schematically described as
 \begin{equation}
 S + E \xrightleftharpoons[k_{-1}]{k_1} C  \xrightharpoonup[]{k_2}    P+ E,
 \label{eq:mm_det}
 \end{equation}
where $k_1$ and $k_{-1}$ are the reaction rate constants for the reversible enzyme binding in the units of $M^{-1}s^{-1}$ and $s^{-1}$ while $k_2$ is the rate constant for the product creation in the unit of $s^{-1}$. Applying the law of mass action to (\ref{eq:mm_det}), temporal changes of the concentrations are described by the following system of ODEs:
 \begin{equation}
 \label{det_eq}
 \begin{split}
 \frac{d[S]}{dt} &= -k_1 [S][E] + k_{-1}[C], \\
 \frac{d[E]}{dt} &= -k_1 [S][E] + k_{-1}[C] + k_2 [C], \\
 \frac{d[C]}{dt} &= k_1 [S][E] - k_{-1}[C] - k_2 [C],\\
 \frac{d[P]}{dt} &= k_2 [C],
 \end{split}
 \end{equation}
 where  the bracket notation $[\cdot]$ refers  to   the concentration of   species. In a closed system, there are two conservation laws for the total amount of enzyme and substrate
 \begin{equation}
 \label{total_cont}
 \begin{split}
 [E_0] &\equiv [E] + [C],\\
 [S_0] &\equiv [S] + [C] + [P].
 \end{split}
 \end{equation}
 These conservation laws not only reduce (\ref{det_eq}) to two equations, but also play an
 important role in the analysis of the reaction network given in (\ref{eq:mm_det}). It is worth mentioning that some authors also consider an additional reversible reaction in the form of binding of the product ($P$) and the free enzyme ($E$) to produce the enzyme-substrate complex ($C$), i.e., $P+ E  \xrightharpoonup[]{} C   $.
 We remark that should we  expand the model in (\ref{eq:mm_det}) to include such a reaction, our discussion in later sections would remain largely  the same requiring  only    simple modifications.

 Leonor Michaelis and Maud Menten investigated the enzymatic kinetics in (\ref{eq:mm_det}) and proposed a
 mathematical model for it in \cite{Michaelis:1913:DKD}.  They
 suggested an  approximate solution for the initial velocity of the enzyme inversion reaction in terms of the substrate concentrations.
 Following their work, numerous attempts have been made to obtain approximate solutions of (\ref{det_eq}) under various quasi-steady-state assumptions. Several conditions on the rate constants have also been proposed for the validity of such  approximations. For example, Briggs and Haldane mathematically derived the Michaelis-Menten equation, which is now known as sQSSA~\cite{Briggs:1925:NKE}. The sQSSA is based on the assumption that the complex reaches its steady state quickly after a transient time, i.e.,  $d[C]/dt\approx 0$ \cite{Segel:1989:QSS}. This approximation is found to be inaccurate when the enzyme concentration is large compared to that of the substrate. The condition for the validity of the sQSSA was first suggested as $[E_0]\ll [S_0]$ by Laidler \cite{Laidler:1955:TTP}, and a more general condition was derived as $[E_0]\ll [S_0]+K_M$ by Segal \cite{Segel:1988:VSS} and Segel and Slemrod \cite{Segel:1989:QSS}, where $K_M\equiv (k_2+k_{-1})/k_1$ is the  so-called Michaelis-Menten constant.

 Borghans et al. later extended the sQSSA to the case with an excessive amount of enzyme and derived the tQSSA by introducing a new variable for total substrate concentration~\cite{Borghans:1996:EQS}. In the tQSSA, one assumes that the total substrate concentration changes on a slow time scale and that the complex reaches its steady state quickly after a transient time, $d[C]/dt\approx 0$. Then, the complex concentration $[C]$ is found as  a solution of a quadratic equation. Approximating $[C]$ in a simple way, they proposed a necessary and sufficient condition for the validity of  tQSSA as
 \begin{equation}
 \left([E_0]+[S_0]+K_M\right)^2 \gg K [E_0], \label{tQSSA_con_intro}
 \end{equation}
 where $K=k_2/k_1$ is the so-called Van Slyke-Cullen constant~\cite{VanSlyke:1914:MAU}. Later, Tzafriri~\cite{Tzafriri:2003:MMK}
 revisited the tQSSA and derived another set of  sufficient conditions for the validity of
 the tQSSA as
 $\epsilon \equiv \left(K/(2[S_0])\right) f\left(r([S_0])\right)\ll 1$
 where $f(r)=(1-r)^{-1/2}-1$ and $r([S_0])=4[E_0][S_0]/\left([E_0]+[S_0]+K_M\right)^2$. He argued that this sufficient condition was always roughly satisfied by showing $\epsilon$ was less than $1/4$ for all values of $[E_0]$ and $[S_0]$. The tQSSA was later improved by
 Dell'Acqua and Bersani~\cite{DellAcqua:2012:PSM} at high enzyme concentrations when (\ref{tQSSA_con_intro}) is satisfied.

 The rQSSA was first suggested as an alternative to  the sQSSA by Segel and Slemrod~\cite{Segel:1989:QSS}. In the rQSSA, the
 substrate, instead of the complex,  was assumed to be at  steady state, $d[S]/dt\approx 0$, and the domain of the validity of the rQSSA was suggested as $[E_0]\gg K$. Then, Schnell and Maini showed that at high enzyme concentration, the assumption $d[S]/dt\approx 0$ was more appropriate in the rQSSA than the assumption $d[C]/dt\approx 0$ used in the sQSSA or tQSSA due to possibly large error during the initial stage of the reactions~\cite{Schnell:2000:EKH}. They derived  necessary conditions for the validity of the rQSSA as $[E_0]\gg K$ and $[E_0]\gg [S_0]$. In the following sections, we will provide alternative derivations of theses different conditions.

\setcounter{equation}{0}
\section{Multiscale stochastic Michaelis-Menten kinetics}
\label{sec:mm_approximation}

Let $X_S$, $X_E$, $X_C$, and $X_P$ denote the copy numbers of molecules of the substrates ($S$), the enzymes ($E$), the enzyme-substrate complex ($C$), and the product ($P$) respectively. We assume the evolution of these copy numbers is governed by a Markovian dynamics given by the following stochastic equations:
\begin{equation}
\label{eq:unscaled_mm}
\begin{split}
X_S(t) &= X_S(0)  -Y_1\left(\int_0^t\kappa_1'X_S(s)X_E(s)\,ds\right) + Y_{-1}\left(\int_0^t\kappa_{-1}'X_C(s)\,ds\right), \\
X_E(t) &= X_E(0)  - Y_1\left(\int_0^t\kappa_1'X_S(s)X_E(s)\,ds\right) + Y_{-1}\left(\int_0^t\kappa_{-1}'X_C(s)\,ds\right) + Y_2\left(\int_0^t\kappa_2'X_C(s)\,ds\right), \\
X_C(t) &= X_C(0) + Y_1\left(\int_0^t\kappa_1'X_S(s)X_E(s)\,ds\right) - Y_{-1}\left(\int_0^t\kappa_{-1}'X_C(s)\,ds\right) -Y_2\left(\int_0^t\kappa_2'X_C(s)\,ds\right), \\
X_P(t) &= X_P(0) + Y_2\left(\int_0^t\kappa_2'X_C(s)\,ds\right) ,
\end{split}
\end{equation}
where $Y_1, Y_{-1}$ and $Y_2$ are independent unit Poisson processes and $t\ge0$.
We denote  $X_{E_0}\equiv X_E(t)+X_C(t)$ and $X_{S_0}\equiv X_S(t)+X_C(t)+X_P(t)$,
and  as in the deterministic model   \eqref{det_eq} in  previous section  assume that the total substrate and enzymes copy numbers, $X_{S_0} $ and $X_{E_0}$, are conserved in time. As shown in \cite{Ball:2006:AAM,Kang:2013:STM}, the representation (\ref{eq:unscaled_mm}) is especially helpful in analyzing systems with multiple time scales or involving species with abundances varying over  different orders of magnitude.  Unlike the chemical master equations, (\ref{eq:unscaled_mm}) explicitly reveals the relations between the species abundances  and the reaction rates.

In the reaction system (\ref{eq:mm_det}), various scales can exist in the species numbers and reaction rate constants, which determine time scales of the species involved. In order to relate these scales, we first define a scaling parameter $N$ to express the orders of magnitude of species copy numbers and rate constants as powers of $N$. We note that $1/N$ plays a similar role as  $\epsilon$ in the singular perturbation analysis of deterministic models~\cite{Segel:1989:QSS}. Denoting scaling exponents for the species $i$ and the $k$th rate constant by $\alpha_i$ and $\beta_k$ respectively, we express unscaled species copy numbers and rate constants as some powers of $N$ as
\begin{equation}
	\begin{split}
	X_i(t) = N^{\alpha_i} Z_i^{N}(t),\,\,\mbox{for $i=S,E,C,P$} \\
	\text{and } \kappa_k' = N^{\beta_k} \kappa_k,\,\,\mbox{for $k=1,-1,2,$}
	\end{split}
\label{scaling_exponent}
\end{equation}
so that the scaled  variables and  constants, $Z_i^{N}(t)$ and $\kappa_k$, are approximately of order $1$ (denoted as $O(1)$). In $Z_i^{N}$, the superscript represents the dependence of the scaled species numbers on $N$. To express different time scales as powers of $N$, we apply a time change by replacing $t$ with $N^\gamma t$. The scaled species number after the time change is given by
\begin{equation*}
X_i(N^{\gamma} t) = N^{\alpha_i} Z_i^{N}(N^{\gamma}t) \equiv  N^{\alpha_i} Z_i^{N,\gamma} (t).
\end{equation*}
Applying the change of variables,  $\{Z^{N,\gamma}\}\equiv\left\{\left(Z_S^{N,\gamma},Z_E^{N,\gamma},Z_C^{N,\gamma},Z_P^{N,\gamma}\right)\right\}$ becomes a parametrized family of  stochastic processes satisfying
\begin{align}
\label{eq:scaled_mm}
\begin{split}
Z_S^{N,\gamma}(t) ={} & Z_S^N(0)  + N^{-\alpha_S} \left[ - Y_1\left(\int_0^t N^{\rho_1+\gamma} \kappa_1Z_S^{N,\gamma}(s)Z_E^{N,\gamma}(s)\,ds\right) + Y_{-1}\left(\int_0^t N^{\rho_{-1}+\gamma} \kappa_{-1}Z_C^{N,\gamma}(s)\,ds\right) \right], \\
Z_E^{N,\gamma}(t)  ={}& Z_E^N(0)  + N^{-\alpha_E} \left[ - Y_1\left(\int_0^t N^{\rho_1+\gamma} \kappa_1Z_S^{N,\gamma}(s)Z_E^{N,\gamma}(s)\,ds\right) + Y_{-1}\left(\int_0^t N^{\rho_{-1}+\gamma} \kappa_{-1}Z_C^{N,\gamma}(s)\,ds\right)\right. \\
&  {}+ \left. Y_2\left(\int_0^t N^{\rho_2+\gamma} \kappa_2Z_C^{N,\gamma}(s)\,ds\right)\right], \\
Z_C^{N,\gamma}(t) = {}& Z_C^N(0) + N^{-\alpha_C} \left[ Y_1\left(\int_0^t N^{\rho_1+\gamma} \kappa_1Z_S^{N,\gamma}(s)Z_E^{N,\gamma}(s)\,ds\right) -  Y_{-1}\left(\int_0^t N^{\rho_{-1}+\gamma} \kappa_{-1}Z_C^{N,\gamma}(s)\,ds\right) \right. \\
&  { }   - \left. Y_2\left(\int_0^t N^{\rho_2+\gamma} \kappa_2Z_C^{N,\gamma}(s)\,ds\right) \right] ,\\
Z_P^{N,\gamma}(t)  = {}& Z_P^N(0) + N^{-\alpha_P} Y_2\left(\int_0^t N^{\rho_2+\gamma} \kappa_2Z_C^{N,\gamma}(s)\,ds\right),
\end{split}
\end{align}
where $\rho_1 \equiv \alpha_S+\alpha_E+\beta_1$, $\rho_{-1} \equiv   \alpha_C+\beta_{-1}$, and $\rho_2  \equiv  \alpha_C+\beta_2$.  As seen from  (\ref{eq:scaled_mm}), the values of $\rho$'s, $\alpha$'s and $\gamma$'s   determine the temporal dynamics of the scaled random processes. For example, consider the limiting behavior of the scaled process for the first reaction in the equation for $S$,
\begin{equation}
N^{-\alpha_S} Y_1\left(\int_0^t N^{\rho_1+\gamma} \kappa_1Z_S^{N,\gamma}(s)Z_E^{N,\gamma}(s)\,ds\right). \label{normalized_reaction1}
\end{equation}
Assuming that $Z_S^{N,\gamma}$ and $Z_E^{N,\gamma}$ are $O(1)$ in the time scale of interest, the limiting behavior of the scaled process depends upon $\rho_1$, $\alpha_S$, and $\gamma$. If the $\rho_1+\gamma<\alpha_S$, the scaled process converges to zero as $N$ goes to infinity. This means that the number of occurrences of the first reaction is outweighed by the order of magnitude of the species copy number for $S$. When $\rho_1+\gamma=\alpha_S$, the number of occurrences of the first reaction is comparable to the order of magnitude of the species copy number for $S$. Then, using the law of large numbers for the Poisson processes\footnote{The strong law of large numbers states that, for a unit Poisson process~$Y$,  $\frac{1}{N} Y(Nu ) \rightarrow u $ almost surely as $N \rightarrow \infty $, (see~\cite{Ethier:1986:MPC}). }, the limiting behavior of (\ref{normalized_reaction1}) is approximately the same as that of
\begin{equation}
\int_0^t \kappa_1Z_S^{N,\gamma}(s)Z_E^{N,\gamma}(s)\,ds. \label{limit_reation1}
\end{equation}
Lastly, when $\rho_1+\gamma>\alpha_S$, the first reaction occurs so frequently that the scaled process in (\ref{normalized_reaction1}) tends to infinity. The limiting behaviors of other scaled processes are determined similarly. Using the scaled processes involving the reactions where $S$ is produced or consumed, we can choose $\gamma$ so that $Z_S^{N,\gamma}(t)$ becomes $O(1)$. Therefore, we have $\alpha_S=\max(\rho_1+\gamma,\rho_{-1}+\gamma)$, and the \emph{time scale} of $S$ is given by
\begin{equation}
\gamma=\alpha_S-\max(\rho_1,\rho_{-1}). \label{time_scale_S}
\end{equation}
Therefore, the time scales of the species numbers and their limiting behaviors are decided by the scaling exponents for species numbers and reactions, that is, they are dictated by the choice of $\alpha$'s and $\beta$'s.

In order to prevent  the system from vanishing to zero or exploding to infinity in the scaling limit, the parameters $\alpha$'s and $\beta$'s must satisfy what are known as the balance conditions~\cite{Kang:2013:STM}. Essentially,  these conditions ensure that the scaling limit is $O(1)$. Intuitively, the largest order of magnitude of the production of species $i$ should be the same as that  of consumption of species $i$. For instance, in the Michaelis-Menten reaction network described in Section \ref{sec:mm_qssa}, balance for the substrate $S$ can be achieved in two ways. First, through the equation $\rho_1=\rho_{-1}$, which \emph{balances} the binding and unbinding of the enzyme to the substrate; and second, by making $\alpha_S$ large enough so that the imbalance between the occurrences of the reversible binding of the enzyme to substrate can be nullified. This gives a restriction on the time scale $\gamma$ as $\gamma+\max(\rho_1,\rho_{-1})\le \alpha_S$. Combining the equality and inequality for each species, we get species balance conditions as
\begin{equation}
\label{species_balance_conditions}
\begin{split}
\rho_1=\rho_{-1}&\quad\mbox{or}\quad \gamma\le \alpha_S-\max(\rho_1,\rho_{-1}), \\
\rho_1=\max(\rho_{-1},\rho_2)&\quad\mbox{or}\quad \gamma\le \alpha_E-\max(\rho_1,\rho_{-1},\rho_2), \\
\rho_1=\max(\rho_{-1},\rho_2)&\quad\mbox{or}\quad \gamma\le \alpha_C-\max(\rho_1,\rho_{-1},\rho_2), \\
\rho_2+\gamma =0&\quad\mbox{or}\quad \gamma\le \alpha_P-\rho_2.
\end{split}
\end{equation}
Even with conditions (\ref{species_balance_conditions})   satisfied, additional conditions are often required to make the scaled species numbers asymptotically $O(1)$. For each linear combination of species, the collective production and consumption rates should be balanced. Otherwise, the time scale of the new variable consisting of the linear combination of the scaled species will be restricted up to some time. The additional conditions are
\begin{equation}
\label{collective_balance_conditions}
\begin{split}
\rho_2+\gamma =0&\quad\mbox{or}\quad \gamma\le \max(\alpha_S,\alpha_C)-\rho_2, \\
\rho_1=\rho_{-1}&\quad\mbox{or}\quad \gamma\le \max(\alpha_C,\alpha_P)-\max(\rho_1,\rho_{-1}),
\end{split}
\end{equation}
which are obtained by comparing collective production and consumption rates of $S+C$ and $C+P$, respectively.

In the following sections, we use a stochastic model of the Michaelis-Menten kinetics (\ref{eq:unscaled_mm}) and derive the deterministic quasi-steady-state approximate models by applying the multiscale approximations with different scaling subject to \eqref{species_balance_conditions} and \eqref{collective_balance_conditions}.

\setcounter{equation}{0}

\section{Standard quasi-steady-state approximation (sQSSA)} \label{sec:sQSSA}

In the deterministic sQSSA, one assumes that the substrate-enzyme complex $C$ reaches its steady-state quickly after a brief transient phase while the other species are still in their transient states. Therefore, by setting $d[C]/dt\approx 0$, one approximates the steady state concentration of the complex. The steady state equation of the complex in (\ref{det_eq}) and the conservation of the total enzyme concentration in (\ref{total_cont}) give
\begin{eqnarray}
[C] &=& \frac{[E_0][S]}{K_M+[S]}, \label{sQSSA_C}
\end{eqnarray}
where $K_M=(k_{-1}+k_2)/k_1$. The substrate concentration is then given by
\begin{eqnarray}
\frac{d[S]}{dt} &=& -\frac{k_2[E_0][S]}{K_M+[S]}. \label{sQSSA_S}
\end{eqnarray}
The corresponding differential equations for $[E]$ and $[P]$ can be written similarly. This approximation is known as the  sQSSA of the Michaelis-Menten kinetics (\ref{eq:mm_det}) under the deterministic setting.

Now, we use stochastic equations for the species copy numbers in (\ref{eq:unscaled_mm}) and apply the multiscale approximation to derive an analogue of (\ref{sQSSA_C})-(\ref{sQSSA_S}). Equations like (\ref{sQSSA_S}) have been previously derived from the stochastic reaction network~\cite{Darden:1979:PAS,Darden:1982:EKS}. It was also revisited specifically using the multiscale approximation method in \cite{Anderson:2011:CTM,Kang:2013:STM}. However, for the sake of completeness, we furnish a brief description below.  Assuming that $E$ and $C$ are on the faster time scale than $S$ and $P$, consider the following scaled processes:
\begin{equation}
\label{sQSSA_scaling}
\begin{split}
& Z_S^{N,\gamma}(t) = \frac{X_S(N^{\gamma}t)}{N}, \quad Z_P^{N,\gamma}(t) = \frac{X_P(N^{\gamma}t)}{N}, \quad
 Z_E^{N,\gamma}(t) = X_E(N^{\gamma}t), \quad Z_C^{N,\gamma}(t) = X_C(N^{\gamma}t), \\
& \kappa_1' = \kappa_1,\quad \kappa_{-1}' = N \kappa_{-1},\quad \kappa_2' = N \kappa_2,
\end{split}
\end{equation}
that is,
\begin{equation}
\label{sQSSA_exponents}
\begin{split}
& \alpha_S=\alpha_P=1,\qquad \alpha_E=\alpha_C=0,\\
& \beta_1=0,\qquad \beta_{-1}=\beta_2=1.
\end{split}
\end{equation}
We are interested in the time scale of $S$  given in (\ref{time_scale_S}). Plugging in the scaling exponent values in (\ref{sQSSA_exponents}), the time scale of $S$ we are interested in corresponds to $\gamma=0$. Setting $\gamma=0$ in the scaled stochastic equations in (\ref{eq:scaled_mm}) and writing $Z_i^N$ instead of $Z_i^{N,\gamma}$ for $i=S,E,C,P$ one obtains from (\ref{sQSSA_exponents})
\begin{equation}
\label{scaled_sQSSA}
\begin{split}
Z_S^N(t) = {}& Z_S^N(0)  - \frac{1}{N} Y_1\left(\int_0^t N\kappa_1Z_S^N(s)Z_E^N(s)\,ds\right) + \frac{1}{N} Y_{-1}\left(\int_0^t N\kappa_{-1}Z_C^N(s)\,ds\right), \\
Z_E^N(t) = {}&  Z_E^N(0)  - Y_1\left(\int_0^t N\kappa_1Z_S^N(s)Z_E^N(s)\,ds\right) + Y_{-1}\left(\int_0^t N\kappa_{-1}Z_C^N(s)\,ds\right) \\
 & {} + Y_2\left(\int_0^t N\kappa_2Z_C^N(s)\,ds\right), \\
Z_C^N(t) = {} & Z_C^N(0) + Y_1\left(\int_0^t N\kappa_1Z_S^N(s)Z_E^N(s)\,ds\right) - Y_{-1}\left(\int_0^t N\kappa_{-1}Z_C^N(s)\,ds\right) \\
 & {} - Y_2\left(\int_0^t N\kappa_2Z_C^N(s)\,ds\right), \\
Z_P^N(t) = {} & Z_P^N(0) + \frac{1}{N} Y_2\left(\int_0^t N\kappa_2Z_C^N(s)\,ds\right).
\end{split}
\end{equation}
Define  $M\equiv Z_E^N(t)+Z_C^N(t)$ and
\begin{eqnarray*}
\mathbb{Z}_C^N(t) &\equiv& \int_0^t Z_C^N(s)\,ds = Mt - \int_0^t Z_E^N(s)\,ds.
\end{eqnarray*}
Note that $M = Z_E^N(0)+Z_C^N(0)=X_E(0)+X_C(0)$, and that $M$ does not depend on the scaling parameter~$N$. As done in \cite{Anderson:2011:CTM,Kang:2013:STM}, assume that $Z_S^N(0)\to Z_S(0)$. The scaled variables $Z_S^N$ and $\mathbb{Z}_C^N$ are bounded so they are relatively compact in the finite time interval $[0,\mathcal{T}]$,  where $0<\mathcal{T}<\infty$. Then, $\left(Z_S^N,\mathbb{Z}_C^N\right)$ converges to $\left(Z_S,\mathbb{Z}_C\right)$ as $N\to\infty$ and satisfies for every $t>0$,
\begin{eqnarray}
Z_S(t) &=& Z_S(0) - \int_0^t \kappa_1 Z_S(s)\left( M - \dot{\mathbb{Z}}_C(s) \right)\,ds + \int_0^t \kappa_{-1} \dot{\mathbb{Z}}_C(s) \,ds, \nonumber\\
0 &=& \int_0^t \kappa_1 Z_S(s)\left( M - \dot{\mathbb{Z}}_C(s) \right)\,ds - \int_0^t \left(\kappa_{-1}+\kappa_2\right)\dot{\mathbb{Z}}_C(s) \,ds. \label{ssC}
\end{eqnarray}
Note that we get (\ref{ssC}) by dividing the equation for $Z_C^N(t)$ in (\ref{scaled_sQSSA}) by $N$ and taking the limit as $N\to\infty$. From (\ref{ssC}), we get
\begin{equation}
\label{stoch_sQSSA_S}
\begin{split}
\dot{Z}_S(t) &= -\frac{\kappa_2 M Z_S(t)}{\kappa_M + Z_S(t)},\\
\dot{\mathbb{Z}}_C(t) &= \frac{M Z_S(t)}{\kappa_M+Z_S(t)},
\end{split}
\end{equation}
where $\kappa_M=(\kappa_{-1}+\kappa_2)/\kappa_1$, which is precisely the sQSSA.

\begin{figure}[htbp]
\centering
\includegraphics[width=0.65\columnwidth]{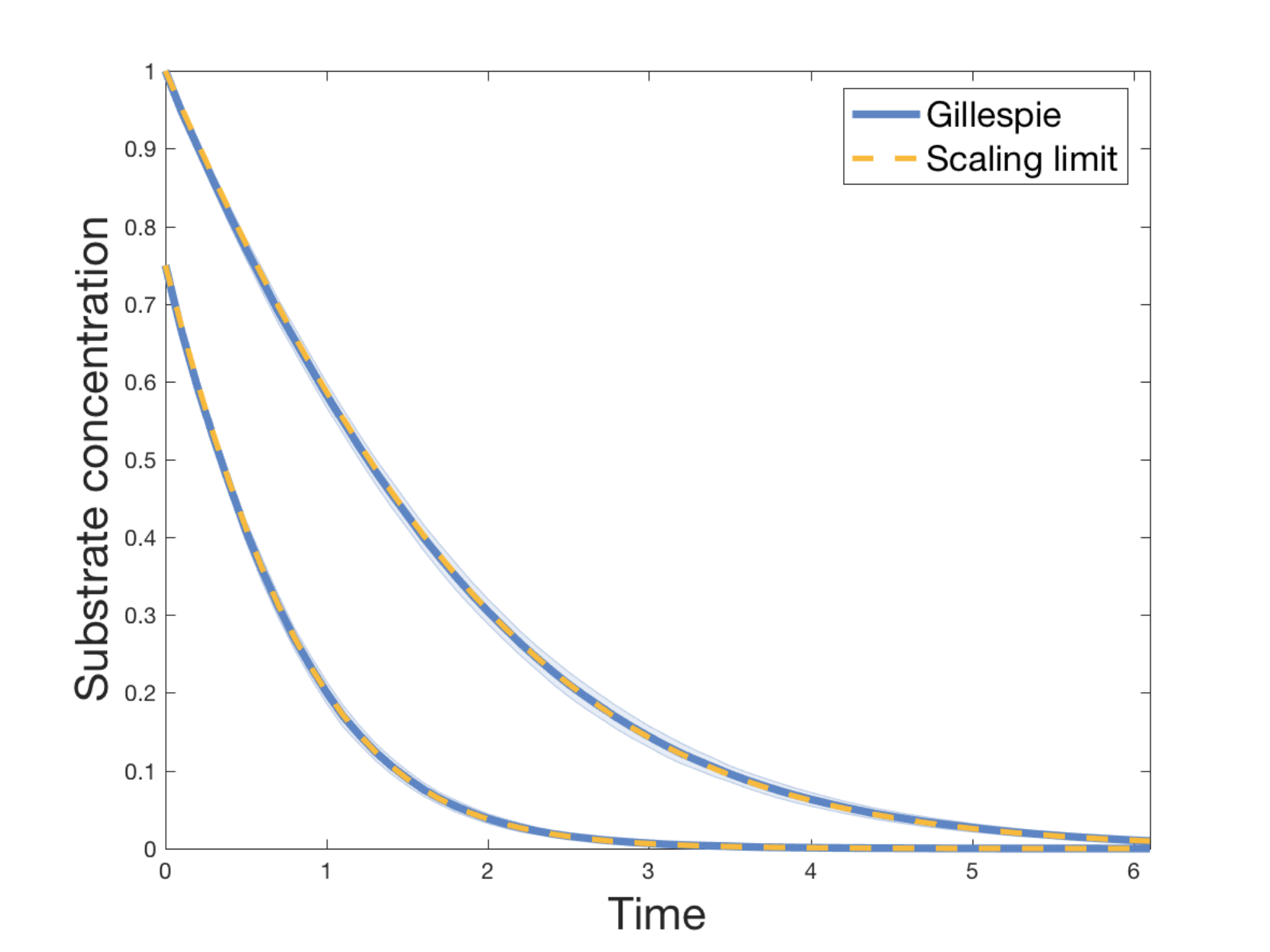}
\caption{\label{fig_sqssa}
Michaelis-Menten kinetics with sQSSA. The limit of the scale substrate copy number, $Z_S(t)$ in (\ref{stoch_sQSSA_S}), is drawn in orange dotted line and the scaled substrate copy number, $Z_S^{N}(t)$ in (\ref{scaled_sQSSA}), is expressed in blue dotted line. The light blue shaded region represents one standard deviation of $Z_S^{N}(t)$ from the mean. The parameter values, $N=1000$, and $(\kappa_1,\kappa_{-1},\kappa_2)=(1,1,0.1)$, and the initial conditions,  $(Z_S^{N}(0), Z_E^{N}(0), Z_C^{N}(0), Z_P^{N}(0))=(1,10,0,0)$,  $M=10$, and $(0.75,20,0,0)$,  $M=20$, respectively for the upper and the lower curves, are used. Two different choices of initial conditions are used to reflect the fact that convergence can be achieved under varying values of the conservation constant~$M$.
}
\end{figure}

Note that we only use a law of large numbers and the conservation law to derive (\ref{stoch_sQSSA_S}). In Figure~\ref{fig_sqssa}, we compare the limit $Z_S(t)$ in (\ref{stoch_sQSSA_S}) with the scaled substrate copy number $Z_S^{N}(t)$ in (\ref{scaled_sQSSA}), obtained from $1000$ realizations of the stochastic simulation using Gillespie's algorithm~\cite{Gillespie:1977:ESS}. Figure~\ref{fig_sqssa} shows  the agreement between the scaled process $Z_S^N(t)$ and its limit $Z_S(t)$.

\paragraph{Conditions for  sQSSA in the deterministic system.}
We have shown that the  scaling exponents (\ref{sQSSA_exponents})  indeed yielded the sQSSA. We now show  how the conditions (\ref{sQSSA_exponents}) are related to the conditions proposed in the literature for the validity of the deterministic sQSSA. First, we consider a general condition derived by Segal \cite{Segel:1988:VSS} and Segel and Slemrod \cite{Segel:1989:QSS},
\begin{eqnarray}
[E_0] &\ll&  [S_0] + K_M, \label{Segel_con}
\end{eqnarray}
where $K_M= (k_{-1}+k_2)/k_1$ is the Michaelis-Menten constant. We rewrite (\ref{Segel_con}) in terms of the species copy numbers and the stochastic reaction rate constants. The stochastic and the deterministic reaction rates are related as
\begin{eqnarray}
\left(k_1, k_{-1}, k_2\right) &=& \left(V \kappa_1', \kappa_{-1}', \kappa_2' \right) \label{rates_rel} ,
\end{eqnarray}
where $V$ is the system volume multiplied by the Avogadro's number~\cite{Kurtz:1972:RSD}. We also use the relation between molecular numbers and molecular concentrations as
\begin{eqnarray}
[i] &=& X_i(t)/V, \quad i=S,E,C,P. \label{variables_rel}
\end{eqnarray}
Applying (\ref{rates_rel}) and (\ref{variables_rel}) in (\ref{Segel_con}), and canceling out $V$, we get
\begin{eqnarray}
X_{E_0} &\ll& X_{S_0}+\frac{\kappa_{-1}' + \kappa_2'}{\kappa_1'}. \label{Segel_con2}
\end{eqnarray}
Plugging our choice of the scaled variables and rate constants given in (\ref{sQSSA_scaling}) in (\ref{Segel_con2}) gives
\begin{eqnarray}
Z_E^N(t) + Z_C^N(t) &\ll& N \left(Z_S^N(t)+Z_P^N(t)\right)+ Z_C^N(t)+\frac{N \left(\kappa_{-1} + \kappa_2\right)}{\kappa_1}. \label{Segel_con3}
\end{eqnarray}
Since $Z_i^N(t)\approx O(1)$ and $\kappa_k\approx O(1)$, the left and the right sides of (\ref{Segel_con3}) become of order $1$ and $N$, respectively. We see that our choice of the scaling in the stochastic model is in agreement with
the conditions for the validity of the sQSSA in the deterministic model (\ref{Segel_con}).

Note that the choice of scaling exponents in (\ref{sQSSA_exponents}) is, in general, not unique. We now derive more general conditions on the scaling exponents, $\alpha$'s and $\beta$'s,  leading to  the sQSSA limit (\ref{stoch_sQSSA_S}). Note that for (\ref{stoch_sQSSA_S}) to hold  the time scale of $C$ should be faster than that of $S$,  so that we can obtain  (\ref{ssC}) from the equation of $C$, i.e., \begin{eqnarray}
\alpha_C - \max(\rho_1,\rho_{-1},\rho_2) &<& \alpha_S-\max(\rho_1,\rho_{-1}) \label{sQSSA_stoch_con1},
\end{eqnarray}
which is an analogue of $d[C]/dt\approx 0$.
Moreover,   for  $E$ to  be expressed in terms of $C$ and  retained in the limit,   the species copy number of $C$ has to be  greater than or equal to that of $E$ in the conservation equation of the total enzyme
\begin{eqnarray}
\alpha_E &\le& \alpha_C \label{sQSSA_stoch_con2}.
\end{eqnarray}
Finally, all reaction propensities are of the same order so that all the terms are present in (\ref{stoch_sQSSA_S})
\begin{eqnarray}
\rho_1=\rho_{-1}=\rho_2 \label{sQSSA_stoch_con3}.
\end{eqnarray}
Combining (\ref{sQSSA_stoch_con1}), (\ref{sQSSA_stoch_con2}), and (\ref{sQSSA_stoch_con3}) together, we get the following conditions
\begin{equation}
\label{sQSSA_stoch_con}
\begin{split}
& \alpha_E\le\alpha_C <\alpha_S, \\
& \alpha_S+\beta_1 = \beta_{-1} = \beta_2.
\end{split}
\end{equation}
The second condition in (\ref{sQSSA_stoch_con}) can be rewritten as $\alpha_S=\beta_{-1}-\beta_1=\beta_2-\beta_1$ and so  (\ref{sQSSA_stoch_con}) implies
\begin{equation*}
\begin{split}
& X_{E_0} \ll X_{S_0}, \\
& X_{E_0} \ll \frac{\kappa_{-1}'}{\kappa_1'} \approx \frac{\kappa_2'}{\kappa_1'},
\end{split}
\end{equation*}
which is comparable to the general condition (\ref{Segel_con}) on the deterministic sQSSA.

\setcounter{equation}{0}

\section{Total quasi-steady-state approximation (tQSSA)} \label{sec:tQSSA}

In the deterministic tQSSA, we define the total substrate concentration as $[T]\equiv [S]+[C]$. Assuming that $[T]$ changes on the slow time scale, the equations (\ref{det_eq})-(\ref{total_cont}) give the following reduced model~\cite{Borghans:1996:EQS,Tzafriri:2003:MMK},
\begin{equation}
\label{det:tQSSA_eq}
\begin{split}
\frac{d[T]}{dt} &= -k_2 [C],\\
\frac{d[C]}{dt} &= k_1 \left\{ \left([T]-[C]\right)\left([E_0]-[C]\right) - K_M [C]\right\},
\end{split}
\end{equation}
where $K_M=(k_{-1}+k_2)/k_1$. Assuming that $d[C]/dt \approx 0$ and using $[C]\le [E_0]$, the unique solution is found as the positive root of a quadratic equation
\begin{eqnarray}
[C] &=& \frac{\left([E_0]+K_M+[T]\right)-\sqrt{\left([E_0]+K_M+[T]\right)^2-4[E_0][T]}}{2}, \label{tQSSA_C}
\end{eqnarray}
and the evolution of the total substrate concentration obeys
\begin{eqnarray}
\frac{d[T]}{dt} &=& -k_2 \frac{\left([E_0]+K_M+[T]\right)-\sqrt{\left([E_0]+K_M+[T]\right)^2-4[E_0][T]}}{2}. \label{tQSSA_T}
\end{eqnarray}
The above  approximation  is the tQSSA of the Michaelis-Menten kinetics (\ref{eq:mm_det}) in  the deterministic setting.

Now, consider the stochastic model  (\ref{eq:unscaled_mm}). Our goal is to  apply the multiscale approximation with the appropriate scaling so that we can consider  (\ref{tQSSA_T}) as the limit of the stochastic Michaelis-Menten system \eqref{eq:scaled_mm} as $N\to \infty$ . We assume that $S$, $E$, and $C$ are on the faster time scale than $P$. Our choice of  scaling is
\begin{equation}
\label{tQSSA_scaling}
\begin{split}
& Z_S^{N,\gamma}(t)=\frac{X_S(N^{\gamma}t)}{N}, \quad Z_E^{N,\gamma}(t)=\frac{X_E(N^{\gamma}t)}{N},\quad Z_C^{N,\gamma}(t)=\frac{X_C(N^{\gamma}t)}{N}, \quad Z_P^{N,\gamma}(t)=\frac{X_P(N^{\gamma}t)}{N}, \\
& \kappa_1' = \kappa_1, \quad \kappa_2' = \kappa_2,\quad \kappa_{-1}' = N \kappa_{-1},
\end{split}
\end{equation}
that is,
\begin{equation}
\label{tQSSA_exponents}
\begin{split}
& \alpha_S=\alpha_E=\alpha_C=\alpha_P=1, \\
& \beta_1=\beta_2=0,\quad \beta_{-1}=1.
\end{split}
\end{equation}
We are interested in the stochastic model in the time scale of $T$. Adding unscaled equations for $S$ and $C$ and dividing by $N^{\max(\alpha_S,\alpha_C)}$ from (\ref{eq:scaled_mm}) we have
\begin{eqnarray*}
\frac{ N^{\alpha_S}Z_S^{N,\gamma}(t)+N^{\alpha_C}Z_C^{N,\gamma}(t) }{N^{\max(\alpha_S,\alpha_C)}}
&=& \frac{ N^{\alpha_S}Z_S^{N}(0)+N^{\alpha_C}Z_C^{N}(0) }{N^{\max(\alpha_S,\alpha_C)}}  \\
&& - \frac{1}{N^{\max(\alpha_S,\alpha_C)} }Y_2 \left(\int_0^t N^{\rho_2+\gamma}\kappa_2 Z_C^{N,\gamma}(s)\,ds\right).
\end{eqnarray*}
Thus, the time scale of $T$ is given by
\begin{eqnarray}
\gamma &=& \max(\alpha_S,\alpha_C) -\rho_2. \label{time_scale_T}
\end{eqnarray}
Using (\ref{tQSSA_exponents}) gives $\gamma=0$. For simplicity, we set the time scale exponent as $\gamma=0$ and denote $Z_i^{N,\gamma}$ as $Z_i^N$ for $i=S,E,C,P$ as we did in Section \ref{sec:sQSSA}.

\begin{figure}[t]
\centering
\includegraphics[width=0.65\textwidth]{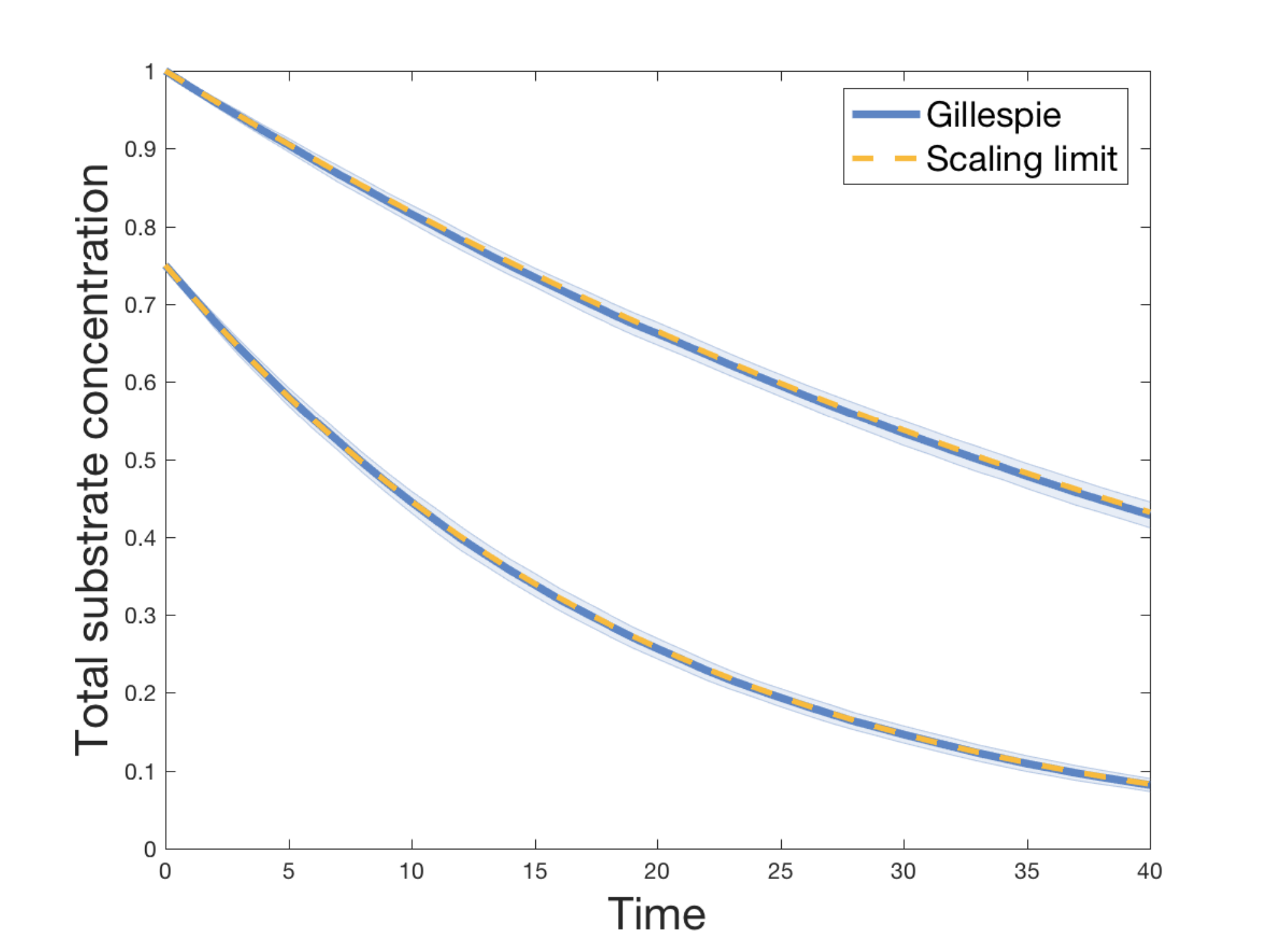}
\caption{\label{fig_tqssa}
Michaelis-Menten kinetics with tQSSA. The limit of the scaled total substrate copy number, $Z_T(t)$ in (\ref{stoch_tQSSA_T}), is drawn in orange dotted line and the scaled total substrate copy number, $Z_T^N(t)$ in (\ref{scaled_tQSSA_T}), is shown in blue. The parameter values, $N=1000$,  and $(\kappa_1,\kappa_{-1},\kappa_2)=(1,4,1)$, and the initial conditions, $(Z_S^N(0), Z_E^N(0), Z_C^N(0), Z_P^N(0))=(1,0.1,0,0)$, $m=0.1$, and $(Z_S^N(0), Z_E^N(0), Z_C^N(0), Z_P^N(0))=(0.75,0.25,0,0)$,  $m=0.25$, respectively for the upper and the lower curves, are used. Two different choices of initial conditions are used here to show convergence under varying values of the conservation constant~$m$.
}
\end{figure}

With the scaling exponents in (\ref{tQSSA_exponents}), the scaled equations in (\ref{eq:scaled_mm}) become
\begin{equation}
\label{scaled_tQSSA}
\begin{split}
Z_S^N(t) =& Z_S^N(0)  - \frac{1}{N} Y_1\left(\int_0^t N^2 \kappa_1Z_S^N(s)Z_E^N(s)\,ds\right) + \frac{1}{N} Y_{-1}\left(\int_0^t N^2 \kappa_{-1}Z_C^N(s)\,ds\right), \\
Z_E^N(t) =& Z_E^N(0)  - \frac{1}{N} Y_1\left(\int_0^t N^2 \kappa_1Z_S^N(s)Z_E^N(s)\,ds\right) + \frac{1}{N} Y_{-1}\left(\int_0^t N^2 \kappa_{-1}Z_C^N(s)\,ds\right) \\
 & {} + \frac{1}{N} Y_2\left(\int_0^t N \kappa_2Z_C^N(s)\,ds\right), \\
Z_C^N(t) = &Z_C^N(0) + \frac{1}{N} Y_1\left(\int_0^t N^2 \kappa_1Z_S^N(s)Z_E^N(s)\,ds\right) - \frac{1}{N} Y_{-1}\left(\int_0^t N^2 \kappa_{-1}Z_C^N(s)\,ds\right) \\
 & - \frac{1}{N} Y_2\left(\int_0^t N \kappa_2Z_C^N(s)\,ds\right), \\
Z_P^N(t) =& Z_P^N(0) + \frac{1}{N} Y_2\left(\int_0^t N \kappa_2Z_C^N(s)\,ds\right).
\end{split}
\end{equation}
Define the new slow variable
\begin{eqnarray*}
Z_T^N(t) &\equiv& Z_S^N(t) + Z_C^N(t),
\end{eqnarray*}
which satisfies
\begin{eqnarray}
Z_T^N(t) &=& Z_T^N(0)  - \frac{1}{N} Y_2\left(\int_0^t N \kappa_2Z_C^N(s)\,ds\right). \label{scaled_tQSSA_T}
\end{eqnarray}
We have two conservation laws for the total amount of substrate and enzyme, $m^N\equiv Z_E^N(t)+Z_C^N(t)$ and $k^N\equiv Z_T^N(t)+Z_P^N(t)$, and we denote their limits as $N\to\infty$ by  $m$ and $k$, respectively. We also define
\begin{eqnarray*}
\mathbb{Z}_C^N(t) &\equiv & \int_0^t Z_C^N(s)\,ds = m^Nt - \int_0^t  Z_E^N(s)\,ds.
\end{eqnarray*}
Since $Z_T^N(t)\le k^N\to k$ and $\mathbb{Z}_C^N(t)\le m^N t\to mt$, $Z_T^N$ and $\mathbb{Z}_C^N$ are bounded,  they are  also  relatively compact in the finite time interval $t\in [0,\mathcal{T}]$ where $0<\mathcal{T}<\infty$. Since the law of large numbers implies that $Z_T^N(0)\to Z_T(0)$ as $N\to\infty$  then  $\left(Z_T^N, \mathbb{Z}_C^N\right)$ (possibly along a subsequence only) converges to $\left(Z_T,\mathbb{Z}_C\right)$ which satisfies
\begin{eqnarray}
Z_T(t) &=& Z_T(0)  - \int_0^t \kappa_2 \dot{\mathbb{Z}}_C(s)\,ds, \nonumber\\
0 &=& \int_0^t \kappa_1\left(Z_T(s)-\dot{\mathbb{Z}}_C(s)\right)\left(m-\dot{\mathbb{Z}}_C(s)\right)\,ds -  \int_0^t \label{Css_tQ}\kappa_{-1}\dot{\mathbb{Z}}_C(s)\,ds.
\end{eqnarray}
Note that (\ref{Css_tQ}) is the limit as $N\to\infty$ when we divide the equation for the scaled variable of $C$ in (\ref{scaled_tQSSA}) by $N$. Hence, we obtain
\begin{eqnarray}
\dot{\mathbb{Z}}_C(t) &=& \frac{\left(m+\kappa_D+Z_T(t) \right) - \sqrt{\left(m+\kappa_D+Z_T(t)\right)^2-4mZ_T(t)}}{2}, \label{stoch_tQSSA_C} \\
\dot{Z}_T(t) &=& -\kappa_2 \frac{\left(m+\kappa_D+Z_T(t)\right)-\sqrt{\left(m+\kappa_D+Z_T(t)\right)^2-4mZ_T(t)}}{2}, \label{stoch_tQSSA_T}
\end{eqnarray}
where $\kappa_D\equiv \kappa_{-1}/\kappa_1$. The equations (\ref{stoch_tQSSA_C}) and (\ref{stoch_tQSSA_T}) are analogous to  (\ref{tQSSA_C}) and (\ref{tQSSA_T}), respectively. Note that we only have $\kappa_D$ in (\ref{stoch_tQSSA_C})-(\ref{stoch_tQSSA_T}) instead of $K_M=(k_{-1}+k_2)/k_1$ in (\ref{tQSSA_C})-(\ref{tQSSA_T}). The reaction rate $\kappa_2$ disappears, since the propensity of the second reaction is of order of $N$, which is slower than the other two reactions whose propensities are of order $N^2$ as shown in (\ref{scaled_tQSSA}). In Figure \ref{fig_tqssa}, we compare the limit $Z_T(t)$ in (\ref{stoch_tQSSA_T}) and the scaled total substrate copy number $Z_T^{N}(t)$ in (\ref{scaled_tQSSA_T}), obtained from $1000$ realizations of the stochastic simulation using Gillespie's algorithm~\cite{Gillespie:1977:ESS}. The plot indicates  close agreement between the scaled process~$Z_T^{N}(t)$ and its proposed limit~$Z_T(t)$.

\paragraph{Conditions for  tQSSA in the deterministic system.}
To derive tQSSA from (\ref{det:tQSSA_eq}), it is assumed that the total substrate concentration changes in the slow time scale and that the complex reaches its steady state quickly after some transient time, that is, $d[C]/dt\approx 0$. The complex concentration $[C]$ is then  found as the nonnegative  solution of  a quadratic equation. As mentioned earlier,    Borghans et al.~\cite{Borghans:1996:EQS} approximated $[C]$ in a form simpler than   the exact solution in (\ref{tQSSA_C}) and found a necessary and sufficient condition for the validity of the tQSSA as
\begin{eqnarray}
K [E_0] &\ll& \left([E_0]+[S_0]+K_M\right)^2, \label{Borghans_con1}
\end{eqnarray}
where $K=k_2/k_1$ and $K_M=(k_{-1}+k_2)/k_1$. The condition (\ref{Borghans_con1}) is equivalent to
\begin{eqnarray}
1&\ll& \left(1+\frac{[E_0]+[S_0]}{K}+\frac{k_{-1}}{k_2}\right) \left(1+\frac{[S_0]+K_M}{[E_0]}\right) \label{Borghans_con4}
\end{eqnarray}
 and is implied by any one of the following
\begin{equation}
\label{Borghans_con5}
\begin{split}
& K \ll [E_0]+[S_0], \\
& k_2 \ll k_{-1}, \\
& [E_0] \ll [S_0] + K_M.
\end{split}
\end{equation}
We convert concentrations and deterministic rate constants to molecular numbers and stochastic rate constants using (\ref{rates_rel})-(\ref{variables_rel}).
After simplification, the condition in (\ref{Borghans_con1}) becomes
\begin{eqnarray}
\frac{\kappa_2'}{\kappa_1'} X_{E_0} &\ll& \left(X_{E_0}+X_{S_0}+\frac{\kappa_{-1}'+\kappa_2'}{\kappa_1'}\right)^2, \label{Borghans_con2}
\end{eqnarray}
by using the same argument  as in  (\ref{Segel_con2}). Plugging our choice of the scaled variables and rate constants as specified  in (\ref{tQSSA_scaling})
 yields
\begin{align*}
 \frac{\kappa_2}{\kappa_1} N\left(Z_E^N(t)+Z_C^N(t)\right) &  \ll
 \left(N\left(Z_E^N(t)+Z_C^N(t)\right)+N\left(Z_S^N(t)+Z_C^N(t)+Z_P^N(t)\right)+\frac{N\kappa_{-1}+\kappa_2}{\kappa_1}\right)^2. \label{Borghans_con3}
\end{align*}
Since in the above expression the term on the left is $O(N)$ and the term on the right is $O(N^2)$, our choice of  scaling in the stochastic model is in agreement with the condition (\ref{Borghans_con1}) for the validity of the tQSSA in the deterministic model.

We may also derive more general  conditions on the scaling exponents, $\alpha$'s and $\beta$'s, which lead to tQSSA limit in (\ref{stoch_tQSSA_T}). To this end note that the time scale of $C$ is faster than that of $T$ so that we can derive an analogue of $d[C]/dt\approx 0$ in (\ref{Css_tQ})
\begin{eqnarray}
\alpha_C- \max(\rho_1,\rho_{-1},\rho_2) &<& \max(\alpha_S,\alpha_C)-\rho_2 \label{tQSSA_stoch_con1}.
\end{eqnarray}
Moreover, the species copy number of $C$ has an order greater than or equal to that of $S$, since   otherwise  $C$ would disappear in the limit of $T$. Similarly, the species copy number of $C$ has an  order greater than or equal to that of $E$ so that the limit for $E$ can be expressed in terms of a conservation constant and $C$. Therefore, we have
\begin{eqnarray}
\max(\alpha_S ,\,\alpha_E)&\le& \alpha_C. \label{tQSSA_stoch_con2}
\end{eqnarray}
Finally, to obtain a quadratic equation with a square root solution in the limit, the enzyme binding reaction rate should be equal to the unbinding reaction rate. That is,
\begin{eqnarray}
\rho_1= \rho_{-1} . \label{tQSSA_stoch_con3}
\end{eqnarray}
Combining (\ref{tQSSA_stoch_con1}), (\ref{tQSSA_stoch_con2}), and (\ref{tQSSA_stoch_con3}), we get the following conditions
\begin{equation}
\label{tQSSA_stoch_con}
\begin{split}
& \max(\alpha_S,\,\alpha_E)\le\alpha_C, \\
& \beta_2< \beta_{-1}=\alpha_C+\beta_1.
\end{split}
\end{equation}
Note that due to $\beta_2<\beta_{-1}$ in (\ref{tQSSA_stoch_con}), we have the discrepancy between $\kappa_D$ in (\ref{stoch_tQSSA_T}) and $K_M$ in (\ref{tQSSA_T}). The condition (\ref{tQSSA_stoch_con}) implies
\begin{equation}
\label{tQSSA_stoch_con2_extra}
\begin{split}
& X_{S_0}\approx X_{E_0}, \\
& \frac{\kappa_2'}{\kappa_1'} \ll \frac{\kappa_{-1}'}{\kappa_1'}\approx X_{E_0},
\end{split}
\end{equation}
which is  consistent with  the condition $k_2\ll k_{-1}$ in (\ref{Borghans_con5}) that was  also suggested  for the stochastic system  tQSSA in~\cite{Barik:2008:SSE}.

\setcounter{equation}{0}

\section{Reverse quasi-steady-state approximation (rQSSA)} \label{sec:rQSSA}

In the deterministic rQSSA, it is  assumed that the enzyme is in high concentration. In this approximation, two time scales are considered. Starting with an initial condition $([S],[E],[C],[P])=([S_0],[E_0],0,0)$ in (\ref{det_eq}), the enzyme concentration is $[E]\approx [E_0]$ during the initial transient phase. Since there is almost no complex during this time, we get an approximate model as
\begin{equation}
\label{det:rQSSA_eq1}
\begin{split}
\frac{d[S]}{dt} &= -k_1 [E_0][S], \\
\frac{d[C]}{dt} &= k_1 [E_0][S].
\end{split}
\end{equation}
After the initial transient phase, the substrate is depleted. Therefore, we assume that $d[S]/dt\approx 0$ in (\ref{det_eq}) and obtain
\begin{eqnarray}
[S] &=& \frac{k_{-1} [C]}{k_1\left([E_0]-[C]\right)}, \label{det:rQSSA_eq3}
\end{eqnarray}
so that  the differential equation for the complex becomes
\begin{equation}
\label{det:rQSSA_eq2}
\begin{split}
\frac{d[C]}{dt} &= -k_2 [C].
\end{split}
\end{equation}
We refer to the approximation of the system  \eqref{det_eq} by  (\ref{det:rQSSA_eq1})-(\ref{det:rQSSA_eq2})  as the rQSSA of the Michaelis-Menten kinetics  in the deterministic setting.

As in the previous sections, let  us   consider the stochastic equations for the Michaelis-Menten kinetics given by  (\ref{eq:unscaled_mm}) and again apply yet another  multiscale approximation with time change, to derive  the rQSSA in (\ref{det:rQSSA_eq1})-(\ref{det:rQSSA_eq2}). We assume that $S$ and $C$ are on   faster time scale than $E$ and $P$. The following scales are chosen
\begin{equation}
\label{rQSSA_scaling}
\begin{split}
& Z_S^{N,\gamma}(t)=\frac{X_S(N^{\gamma}t)}{N}, \quad Z_E^{N,\gamma}(t)=\frac{X_E(N^{\gamma}t)}{N^2}, \quad
 Z_C^{N,\gamma}(t)=\frac{X_C(N^{\gamma}t)}{N}, \quad Z_P^{N,\gamma}(t)=\frac{X_P(N^{\gamma}t)}{N}, \\
& \kappa_1' = \kappa_1,\quad
 \kappa_{-1}' = N \kappa_{-1}, \quad \kappa_2' = N \kappa_2,
\end{split}
\end{equation}
that is,
\begin{equation*}
\begin{split}
& \alpha_S=\alpha_C=\alpha_P=1,\quad \alpha_E=2, \\
& \beta_1= 0,\quad \beta_{-1}= \beta_2=1.
\end{split}
\end{equation*}
Then, the reduced system is obtained from (\ref{eq:scaled_mm}) using (\ref{rQSSA_scaling}) as
\begin{equation}
\label{scaled_rQSSA}
\begin{split}
Z_S^{N,\gamma}(t) &= Z_S^N(0)  - \frac{1}{N}Y_1\left(\int_0^t N^{\gamma+3} \kappa_1Z_S^{N,\gamma}(s)Z_E^{N,\gamma}(s)\,ds\right) + \frac{1}{N}Y_{-1}\left(\int_0^t N^{\gamma+2} \kappa_{-1}Z_C^{N,\gamma}(s)\,ds\right), \\
Z_E^{N,\gamma}(t) &= Z_E^N(0)  - \frac{1}{N^2} Y_1\left(\int_0^t N^{\gamma+3} \kappa_1Z_S^{N,\gamma}(s)Z_E^{N,\gamma}(s)\,ds\right) + \frac{1}{N^2} Y_{-1}\left(\int_0^t N^{\gamma+2} \kappa_{-1}Z_C^{N,\gamma}(s)\,ds\right) \\
 & + \frac{1}{N^2} Y_2\left(\int_0^t N^{\gamma+2} \kappa_2Z_C^{N,\gamma}(s)\,ds\right), \\
Z_C^{N,\gamma}(t) &= Z_C^N(0) + \frac{1}{N} Y_1\left(\int_0^t N^{\gamma+3} \kappa_1Z_S^{N,\gamma}(s)Z_E^{N,\gamma}(s)\,ds\right) - \frac{1}{N} Y_{-1}\left(\int_0^t N^{\gamma+2} \kappa_{-1}Z_C^{N,\gamma}(s)\,ds\right) \\
 & - \frac{1}{N} Y_2\left(\int_0^t N^{\gamma+2} \kappa_2Z_C^{N,\gamma}(s)\,ds\right), \\
Z_P^{N,\gamma}(t) &= Z_P^N(0) + \frac{1}{N} Y_2\left(\int_0^t N^{\gamma+2} \kappa_2Z_C^{N,\gamma}(s)\,ds\right).
\end{split}
\end{equation}
Note that this choice of scaling does not satisfy the balance equations introduced in (\ref{species_balance_conditions}). The inequalities for $S$ and $C$ give $\gamma\le -2$ and those for $E$ and $P$ give $\gamma\le -1$. These conditions suggest the first and the second time scales as $\gamma=-2$ when $S$ and $C$ become $O(1)$ and $\gamma=-1$ when $E$ and $P$ are $O(1)$.
Define the following conservation constants
\begin{equation}
\label{rqssa_conservation}
\begin{split}
m^N &= Z_E^{N,\gamma}(t)+\frac{1}{N}Z_C^{N,\gamma}(t), \\
k^N &= Z_S^{N,\gamma}(t)+Z_C^{N,\gamma}(t)+Z_P^{N,\gamma}(t),
\end{split}
\end{equation}
which  we assume to converge to some limiting values $m$ and $k$ as $N\to\infty$, respectively. In this setting, $Z_S^{N,\gamma}$, $Z_E^{N,\gamma}$, $Z_C^{N,\gamma}$, and $Z_P^{N,\gamma}$ are bounded so that they are relatively compact for  $t\in[0,\mathcal{T}]$, where $0<\mathcal{T}<\infty$. In the first time scale when $\gamma=-2$, the scaled species for $E$ and $P$ converge to their initial conditions, $Z_E^{N,-2}(t)\to Z_E(0)$ and $Z_P^{N,-2}(t)\to Z_P(0)$ as $N\to\infty$, since the scaling exponents in the propensities are greater than those of species copy numbers in this time scale. Therefore $\left(Z_S^{N,-2},Z_C^{N,-2}\right)$ converges to $\left(Z_S^{(-2)},Z_C^{(-2)}\right)$ satisfying
\begin{equation}
\label{stoch_rQSSA_time1}
\begin{split}
Z_S^{(-2)}(t) &= Z_S(0)  - \int_0^t  \kappa_1Z_S^{(-2)}(s)Z_E(0)\,ds,\\
Z_C^{(-2)}(t) &= Z_C(0) + \int_0^t  \kappa_1Z_S^{(-2)}(s)Z_E(0)\,ds.
\end{split}
\end{equation}
Since $Z_C^{N,-2}(t)$ is bounded by $k^N$ from (\ref{rqssa_conservation}), the remaining reaction terms for the unbinding of the complex and for the product production vanish  as $N\to\infty$. The equations \eqref{stoch_rQSSA_time1} are seen as the integral version of \eqref{det:rQSSA_eq1}, that is, the   rQSSA   for the first (transient) time scale.

Next, consider the second time scale when $\gamma=-1$. Plugging $\gamma=-1$ in the equation for $S$ in (\ref{scaled_rQSSA}), and applying the law of large numbers, we obtain
\begin{eqnarray}
Z_S^{N,-1}(t) &\approx& Z_S^N(0) - \int_0^t \left( N \kappa_1 Z_S^{N,-1}(s)Z_E^{N,-1}(s) - \kappa_{-1}Z_C^{N,-1}(s)\right)\,ds. \label{scaled_rQSSA_S}
\end{eqnarray}
Using (\ref{scaled_rQSSA_S}), the equations for $E$ and $C$ in (\ref{scaled_rQSSA}) become
\begin{eqnarray}
Z_C^{N,-1}(t) &\approx& Z_C^N(0) +  Z_S^N(0) - Z_S^{N,-1}(t)
- \int_0^t \kappa_2 Z_C^{N,-1}(s)\,ds,
\label{scaled_rQSSA_C} \\
Z_E^{N,-1}(t) &\approx& Z_E^N(0)  - \int_0^t \kappa_1Z_S^{N,-1}(s)Z_E^{N,-1}(s)\,ds,
\label{scaled_rQSSA_E}
\end{eqnarray}
since  the remaining reaction terms are asymptotically equal to zero.
Dividing (\ref{scaled_rQSSA_S}) by $N$, we obtain
\begin{eqnarray}
\int_0^t \kappa_1 Z_S^{N,-1}(s)Z_E^{N,-1}(s)\,ds &\to& 0, \label{first_reaction}
\end{eqnarray}
as $N\to\infty$, since all other terms vanish  asymptotically. Due to \eqref{scaled_rQSSA_E} and (\ref{first_reaction}), $Z_E^{N,-1}(t)\to Z_E(0)$ as $N\to\infty$.
 Defining $\mathbb{Z}_S^{N,-1}(t)\equiv \int_0^t Z_S^{N,-1}(s)\,ds$ and using (\ref{first_reaction}) and (\ref{scaled_rQSSA_C}), we conclude that $\left(\mathbb{Z}_S^{N,-1},Z_C^{N,-1}\right)$ converges to $\left(\mathbb{Z}_S^{(-1)},Z_C^{(-1)}\right)$ satisfying
\begin{eqnarray}
0 &=& \int_0^t \kappa_1 \dot{\mathbb{Z}}_S^{(-1)}(s) Z_E(0) \,ds, \nonumber\\
Z_C^{(-1)}(t) &=& Z_C(0) + Z_S(0) - \dot{\mathbb{Z}}_S^{(-1)}(t)  - \int_0^t \kappa_2 Z_C^{(-1)}(s)\,ds.\label{rQSSA_limit_C}
\end{eqnarray}
Therefore,
\begin{equation}
\label{stoch_rQSSA_time2}
\begin{split}
\dot{\mathbb{Z}}_S^{(-1)}(t) &= 0, \\
\dot{Z}_C^{(-1)}(t) &= -\kappa_2 Z_C^{(-1)}(t),
\end{split}
\end{equation} which is the analogue of the rQSSA in the second time scale \eqref{det:rQSSA_eq3}-\eqref{det:rQSSA_eq2} as derived from the deterministic model.

\begin{figure}[t]
\centering
\includegraphics[width=0.65\textwidth]{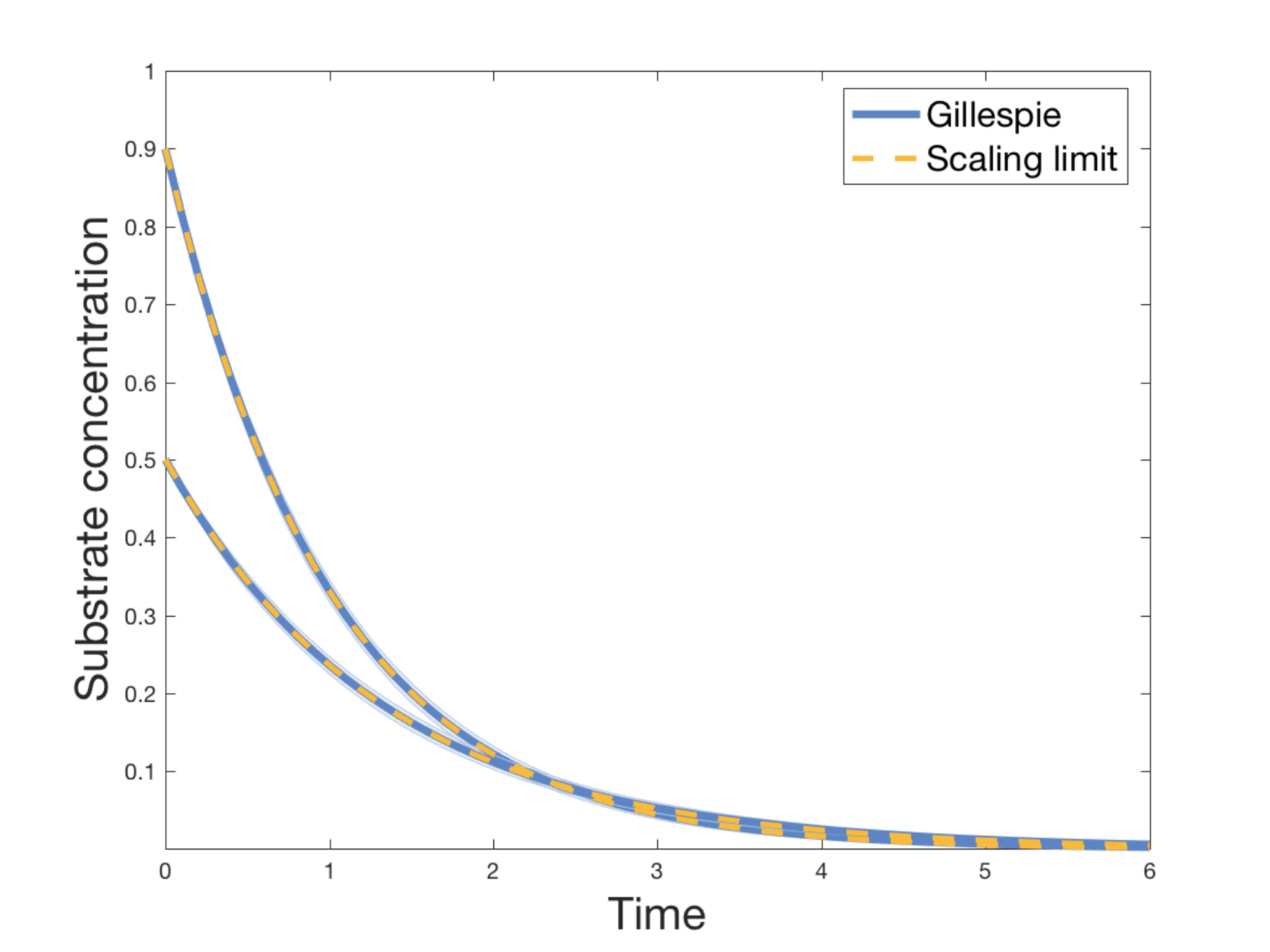}
\caption{\label{fig_rqssa1}
Michaelis-Menten kinetics with rQSSA in the first time scale: The limit of the scaled substrate copy number, $Z_S^{(-2)}(t)$ in (\ref{stoch_rQSSA_time1}), is drawn in orange dotted line and the scaled substrate copy number, $Z_S^{N,-2}(t)$ in (\ref{scaled_rQSSA}), is shown in blue  line. The parameter values, $N=1000$, and $(\kappa_1,\kappa_{-1},\kappa_2)=(1,1,0.1)$, and the initial conditions, $(Z_S^{N}(0), Z_E^{N}(0), Z_C^{N}(0), Z_P^{N}(0))=(0.9,1,0.01,0)$,  $m=1.01$, and $(Z_S^{N}(0), Z_E^{N}(0), Z_C^{N}(0), Z_P^{N}(0))=(0.5,0.75,0.11,0)$,  $m=0.86$, respectively for the upper and the lower curve,  are used. Given the scaling assumptions, the convergence is not sensitive to the exact values of the initial conditions. The only purpose of the two different sets of initial conditions is to illustrate convergence under varying values of the conservation constant~$m$.
}
\end{figure}

\begin{figure}[t]
\centering
\includegraphics[width=0.65\textwidth]{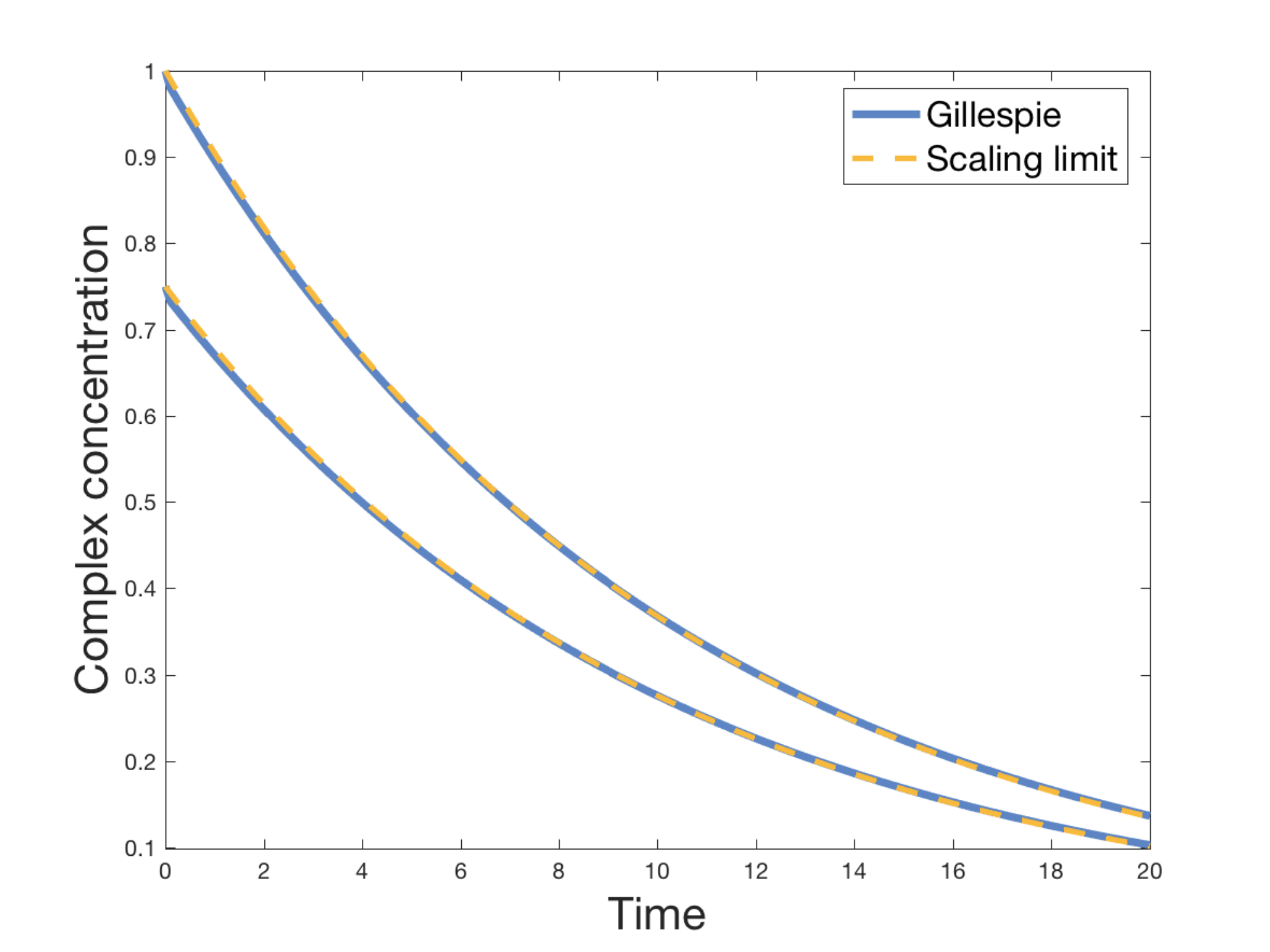}
\caption{\label{fig_rqssa2}
Michaelis-Menten kinetics with rQSSA in the second time scale: The limit of the scaled complex copy number, $Z_C^{(-1)}(t)$ in (\ref{stoch_rQSSA_time2}), is drawn in orange dotted line and the scaled complex copy number, $Z_C^{N,-1}(t)$ in (\ref{scaled_rQSSA}), is shown in blue line. The parameter values, $N=10000$ and $(\kappa_1,\kappa_{-1},\kappa_2)=(1,1,0.1)$, and the initial conditions,  $(Z_S^{N}(0), Z_E^{N}(0), Z_C^{N}(0), Z_P^{N}(0))=(0,0.01,1,0)$, $m=1.01$, and $(Z_S^{N}(0), Z_E^{N}(0), Z_C^{N}(0), Z_P^{N}(0))=(0,0.0075,0.75,0)$, $m=0.7575$ respectively for the upper and the lower curves, are used. Given the scaling assumptions, the convergence is not sensitive to the exact values of the initial conditions. The   two different sets of initial conditions only illustrate convergence under varying values of the conservation constant~$m$.
  }
\end{figure}

We  illustrate  the quality of rQSSA in the stochastic Michaelis-Menten system with some simulations.
In Figure \ref{fig_rqssa1}, we compare the limit $Z_S^{(-2)}(t)$ in (\ref{stoch_rQSSA_time1}) and the scaled substrate copy number $Z_S^{N,-2}(t)$ in \eqref{scaled_rQSSA} using $1000$ runs of the Gillespie's algorithm. In Figure \ref{fig_rqssa2}, we compare the limit $Z_C^{(-1)}(t)$ in (\ref{stoch_rQSSA_time2}) and the scaled complex copy number $Z_C^{N,-1}(t)$ in (\ref{scaled_rQSSA}) using $10000$ runs of the Gillespie's algorithm. Note that the initial condition of $Z_C^{(-1)}(t)$ is $Z_C(0)+Z_S(0)$ in (\ref{rQSSA_limit_C}). However, this does not affect since $Z_S(0)=0$ in our simulation in Figure \ref{fig_rqssa2}.
In both time scales, the scaled processes are in close agreement with the proposed limits.
\paragraph{Conditions for  rQSSA in the deterministic system.}
Consider the general condition for the validity of the rQSSA at high enzyme concentrations suggested by Schnell and Maini~\cite{Schnell:2000:EKH},
\begin{eqnarray}
&& K \ll [E_0] \quad \mbox{and}\quad [S_0] \ll [E_0], \label{Schnell_con}
\end{eqnarray}
where $K=k_2/k_1$.  Rewriting (\ref{Schnell_con}) in terms of molecular copy numbers and stochastic rate constants using (\ref{rates_rel})-(\ref{variables_rel}) gives
\begin{eqnarray}
\frac{\kappa_2'}{\kappa_1'} \ll X_{E_0} \quad \mbox{and}\quad
X_{S_0} \ll X_{E_0}, \label{Schnell_con2}
\end{eqnarray}
since  $V$'s all cancel out. Using our choice of scaling in (\ref{rQSSA_scaling}), the conditions (\ref{Schnell_con2}) become
\begin{equation}
\label{Schnell_con3}
\begin{split}
& \frac{N \kappa_2}{\kappa_1} \ll \left(N^2 Z_E^{N,\gamma}(t)+N Z_C^{N,\gamma}(t)\right) \quad \mbox{and}\quad \\
& N \left(Z_S^{N,\gamma}(t)+Z_C^{N,\gamma}(t)+Z_P^{N,\gamma}(t)\right) \ll \left(N^2 Z_E^{N,\gamma}(t)+N Z_C^{N,\gamma}(t)\right).
\end{split}
\end{equation}
Since the inequalities  in (\ref{Schnell_con3}) hold for large $N$,  our   choice of  scaling is seen  to satisfy  the conditions (\ref{Schnell_con}).

As seen in the previous sections, we may  also derive more general conditions on the scaling exponents, $\alpha$'s and $\beta$'s, leading to (\ref{stoch_rQSSA_time1}) and (\ref{stoch_rQSSA_time2}). In the first scaling, the time scales of $S$ and $C$ are the same and faster than   the time scale of $E$. Therefore it follows  that
\begin{eqnarray}
&& \alpha_S- \max(\rho_1,\rho_{-1}) = \alpha_C- \max(\rho_1,\rho_{-1},\rho_2) < \alpha_E - \max(\rho_1,\rho_{-1},\rho_2). \label{rQSSA_stoch_con1}
\end{eqnarray}
Since the binding reaction rate of the enzyme is faster than the rates of the other two reactions as we see in the limit (\ref{stoch_rQSSA_time1}), we have
\begin{eqnarray}
\max(\rho_{-1},\rho_2) &<& \rho_1. \label{rQSSA_stoch_con2}
\end{eqnarray}
Combining (\ref{rQSSA_stoch_con1}) and (\ref{rQSSA_stoch_con2}), the conditions in the first time scale are
\begin{equation}
\label{rQSSA_stoch_con_time1}
\begin{split}
& \alpha_S=\alpha_C<\alpha_E, \\
& \max(\beta_{-1},\beta_2) < \alpha_E+\beta_1.
\end{split}
\end{equation}
Then, the condition in (\ref{rQSSA_stoch_con_time1}) implies
\begin{equation}
\begin{split}
& X_{S_0} \ll X_{E_0}, \\
& \max\left(\frac{\kappa_{-1}'}{\kappa_1'},\frac{\kappa_2'}{\kappa_1'}\right) \ll X_{E_0},
\end{split}
\end{equation}
which is comparable to   (\ref{Schnell_con}).

Next, consider the second time scale and the condition on the scaling exponents that yields (\ref{stoch_rQSSA_time2}). Note that the conditions (\ref{rQSSA_stoch_con1})-(\ref{rQSSA_stoch_con2}) are already sufficient to derive the limiting process in the second time scale. The condition (\ref{rQSSA_stoch_con1}) implies the time scales of $S$ and $C$ are the same. Since $\rho_2<\rho_1$ as in (\ref{rQSSA_stoch_con2}), the time scale of $S+C$ is slower than that of $S$. Setting the time scale of $S+C$ as the reference one,  we see that on  that timescale  $S$ will be rapidly depleted and then approximated by  zero in view of  the discrepancy between  the consumption and production rates of $S$,  due to $\rho_{-1}<\rho_1$ in (\ref{rQSSA_stoch_con2}). Therefore, the conditions in (\ref{rQSSA_stoch_con1})-(\ref{rQSSA_stoch_con2}) are sufficient to obtain the limit in (\ref{stoch_rQSSA_time2}) on  the second time scale as well. Finally, note that the stochastic Michaelis-Menten system with (\ref{rQSSA_stoch_con_time1}) does not provide an analogue equation for $S$ in (\ref{det:rQSSA_eq3}) due to the condition, $\rho_{-1}<\rho_1$, as shown in (\ref{rQSSA_stoch_con2}). Assuming  $\rho_{-1}=\rho_1$ will  balance production and consumption of $S$, but in this case we can no longer claim  the relative compactness of $S$.

\setcounter{equation}{0}
\section{Discussion}
\label{sec:discussion}

In this paper, we derived the  sQSSA,  the  tQSSA and  the rQSSA for the Michaelis-Menten model of enzyme  kinetics from  general stochastic equations describing  interactions between enzyme, substrate and enzyme-substrate complex in terms of a jump Markov process.   We have shown that these various QSSAs are a consequence of the law of large numbers for the stochastic chemical reaction network under appropriately chosen scaling regimes. Our derivation relies on the multiscale  approximation approach \cite{Kang:2013:STM,Ball:2006:AAM} that  is   quite general and could be  used  to obtain  similar types of QSSAs in  other stochastic chemical reaction systems. One possible  example is  a model of signal transduction into protein phosphorylation cascade, such as  the mitogen-activated protein kinase (MAPK) signaling pathway~\cite{Bersani:2005:MAS,GomezUribe:2007:ORS,DellAcqua:2011:QSS}. In  MAPK signaling pathway, the product of one level of the cascade may act as the enzyme at the  next level, with  different Michaelis-Menten QSSAs found to  be appropriate  at different   levels ~\cite{Sauro:2004:QAS,Bersani:2005:MAS,GomezUribe:2007:ORS,DellAcqua:2011:QSS}.

\begin{table}[tb!]
\begin{center}
\caption{Comparison of  conditions for  the quasi-steady-state approximations in the stochastic  and deterministic Michaelis-Menten kinetics.} \label{table:comparison}
   \begin{tabular}{| l  l  l  l|}  \hline
\multicolumn{1}{|c}{Conditions on} & \multicolumn{1}{c}{sQSSA} & \multicolumn{1}{c}{tQSSA} & \multicolumn{1}{c|}{rQSSA}\\ \hline\hline
stochastic & $\alpha_E\le \alpha_C<\alpha_S$ & $\max(\alpha_S,\,\alpha_E)\le\alpha_C$ & $\alpha_S=\alpha_C<\alpha_E$\\
scaling & $\alpha_S=\beta_{-1}-\beta_1=\beta_2-\beta_1$ & $\beta_2<\beta_{-1}=\alpha_C+\beta_1$ & $\max(\beta_{-1},\beta_2)<\alpha_E+\beta_1$\\
\hline
stochastic  & $X_{E_0}\ll X_{S_0}$ & $X_{E_0}\approx X_{S_0}$ & $X_{S_0}\ll X_{E_0}$\\
abundance & $X_{E_0}\ll \frac{\kappa_{-1}'}{\kappa_1'}\approx \frac{\kappa_2'}{\kappa_1'}$ & $\frac{\kappa_2'}{\kappa_1'}\ll \frac{\kappa_{-1}'}{\kappa_1'} \approx X_{E_0}$ &
$\max\left( \frac{\kappa_{-1}'}{\kappa_1'},\frac{\kappa_2'}{\kappa_1'} \right) \ll X_{E_0}$\\
 \hline
deterministic & $[E_0]\ll [S_0]+K_M$ & $K[E_0]\ll \left([E_0]+[S_0]+K_M\right)^2$ & $K\ll [E_0]$ and $[S_0]\ll [E_0]$\\
abundance & & &\\
   \hline
\end{tabular}
\end{center}
\footnotesize{The parameters are $K=k_2/k_1$ and $K_M=(k_{-1}+k_2)/k_1$.}
\end{table}

Since the dynamics  of  enzyme kinetics plays such a central  role in many problems of modern biochemistry,
 it is important to understand the precise conditions for   the QSSA's discussed here.  For convenience, in Table \ref{table:comparison}, we summarize the conditions for different  QSSAs  in terms of  their  scaling exponents as well as the stochastic and deterministic species abundances.  The  conditions for the stochastic  scalings presented  in the first row of the table clearly separate the range of parameter values intro three regimes.  As we can see, the exponent $\alpha_S$ should be greater than the other exponents for species copy numbers in the sQSSA while $\alpha_E$ is greater than the other exponents for species copy numbers in the rQSSA. In the tQSSA, $\alpha_C$ needs to be greater than or equal to the other exponents.   For the sQSSA and the rQSSA,  the  stochastic  species abundance conditions (listed in the second row) are seen to also  imply the deterministic abundance conditions (listed in the third row).    However, the necessary condition for the tQSSA derived from the stochastic model is slightly different from the corresponding deterministic condition as  it requires the similar order of magnitude for the total amount of enzyme and the total amount of substrate. Note, however,  that the condition  on the deterministic rates  $k_2\ll k_{-1}$, which is an  analog of the stochastic rates condition $\kappa_2'\ll\kappa_{-1}'$, implies  both the deterministic and the stochastic abundance conditions for the  tQSSA.

Our derivations of the QSSAs from the stochastic Michaelis-Menten kinetics provide approximate ODE models where reaction propensities  follow rational or square-root functions and hence violate  the law of mass action.  Such non-standard propensity functions   are often useful  for building  efficient reduced model also in the stochastic settings where they may be used as  intensity  functions in  the random time change representation of the Poisson processes.  For instance, Grima et al.\cite{Grima:2012:SFG}, Chow et al. \cite{Choi:2017:MME}, as well as some  others \cite{Tian:2006:SMR,Kim:2012:SGE}
have  applied this  idea to construct approximate,   stochastic  Michaelis-Menten  enzyme kinetic networks and even the  gene regulatory networks   \cite{Smith:2015:AAS}.  As some of the  authors of this article argued in their recent work (\cite{Kim:2017:RSB}),  such  approximate stochastic models using intensities derived from  the deterministic limits may in some sense be better approximations of   the  underlying stochastic networks than the deterministic QSSAs. Our derivations presented here   could be used to further justify this  statement, at least for  networks satisfying  certain scaling conditions \cite{Rao:2003:SCK,Sanft:2011:LSM,Kim:2015:RSD}, including  those  presented in  Table~1. We  therefore  hope that the results in the  current paper will further contribute to  developing more  accurate approximations of models for enzyme kinetics in biochemical networks.


\section{Acknowledgements}

	This work has been co-funded by the German Research Foundation~(DFG) as part of project C3 within the Collaborative Research Center~(CRC) 1053 -- MAKI (WKB) and the National Science Foundation under the grants RAPID DMS-1513489 (GR) and DMS-1620403 (HWK). This research has also been supported in part by the University of Maryland Baltimore County under grant UMBC KAN3STRT (HWK). This work was initiated when HWK and WKB 
	were visiting the Mathematical Biosciences Institute (MBI) at the Ohio State University in Winter 2016-17. MBI is receiving major funding from the National Science Foundation under the grant DMS-1440386. HWK and WKB 
	acknowledge the hospitality of MBI during their visits to the institute.

\bibliographystyle{unsrt}      
\bibliography{bibqssa.bib}   

\end{document}